\documentclass{article}
\usepackage{latexsym}
\begin{document}
\title{Brane Gravity from Bulk Vector Field}
\author{{\bf Merab Gogberashvili}\\
Andronikashvili Institute of Physics, 6 Tamarashvili Str.,
Tbilisi 380077, Georgia \\
{\sl E-mail: gogber@hotmail.com }}
\maketitle
\begin{abstract}
It is shown that Einstein's equations on the brane can be received from the 
multi-dimensional vector field equations in pseudo-Euclidean space. The idea is 
based on the observation that the brane geometry can be equivalently described 
by the intrinsic metric or by the derivatives of its normal. From the other 
hand the normal to the brane can be constructed with the components of some 
multi-dimensional vector fields. For the both cases 4-dimensional effective 
action for gravity appears to be the same. 

\medskip 

\end{abstract}
\medskip

Since the over 80 years old paper of Theodor Kaluza \cite{Kal} it has been 
believed that Einstein's equations are more fundamental then matter fields 
equations. In standard Kaluza-Klein approach matter fields are considered to 
be a part of multi-dimensional metric tensor. However, the difficulties of 
General Relativity are well known and it is still not clear whether gravity is 
a quantum field or some classical effective interaction. 

Recently, it was shown that Plank's scale $M_{Pl}$ can be constructed with the 
fundamental scale $M$ and the brane width $\epsilon$ \cite{ADD}. On the brane, 
possibly, not only $M_{Pl}$, but Einstein equations can be effective as well. 
In multi dimensions the equations describing gravity can be quite different 
from Einstein's equations.

Note that the first published suggestion about the unification of gravity and 
matter in multi dimensions (which was given by Gunnar Nordstrom \cite{Nor} 
anticipating even General Relativity) was based on vector fields approach. 
Nordstrom started from 5-dimensional Maxwell's equations and imposed also the 
'cilinder condition' (the fields should not depend on the fifth coordinate). He
identified an extra component of vector-potential with the gravity potential 
required in his scalar gravity theory. In some recent papers these extra scalar 
components of multi-dimensional vector-potentials are used to explain Higgs 
mechanism on the brane \cite{higgs}.

In this paper we also use a multi-dimensional vector field to describe gravity 
on the brane, but do not follow the simple approach of Nordstrom. When in 
pseudo-Euclidean space a brane is embedded, multi-dimensional vector fields 
together with the brane geometry can imitate Einstein gravity on the brane. 
Here we don't specify the nature of the brane, for example, it can be the kink 
solution of nonlinear equation of some multi-dimensional scalar field. We just 
show that solutions of 4-dimensional Einstein's equations could be constructed 
with the solutions of multidimensional Maxwell's equations. In this picture 
gravity exhibits tensor character only on the brane and graviton appears to be 
the combination of two 1-spin massless particles. We hope that in such approach 
geometrical unification of different interactions will be easier since Dirac 
equation also can be derived from the constrained Yang-Mills Lagrangian 
\cite{spinor}. Similar ideas inducing gravity (in the linear approximation) on 
some plane in multi dimensions was considered in \cite{Kok} using analogies 
with elasticity theory. 

We want to start with reminding the reader that any $n$-dimensional Riemannian 
space can be embedded into $N$-dimensional pseudo-Euclidean space with 
$n \le N \le n(n+1)/2$ \cite{EF}. Thus, no more than ten dimensions are 
required to embed any 4-dimensional solution of Einstein's equations with 
arbitrary energy-momentum tensor. Embedding the space-time with the 
coordinates $x^\alpha$ and metric $g_{\alpha\beta}$ into pseudo-Euclidean 
space with Cartesian coordinates $\phi^A$ and Minkowskian metric 
$\eta_{AB}$ is given by
\begin{equation}\label{metric} 
ds^{2}= g_{\alpha\beta}dx^\alpha dx^\beta =\eta_{AB}h^A_\alpha h^B_\beta 
dx^\alpha dx^\beta = \eta_{AB}d\phi^Ad\phi^B ~. 
\end{equation} 
Capital Latin letters $A,B,\dots $ labels coordinates of embedded space, 
while Greek indices $\alpha,\beta,\dots$ enumerate coordinates in four 
dimensions. Existence of the embedding (\ref{metric}) demonstrates that the 
multi-dimensional 'tetrad' fields $h^{A}_{\alpha}$ can be expressed as a 
derivatives of some vector 
\begin{equation} \label{tetrad}
h^{A}_{\alpha} = \partial_\alpha \phi^A ~.
\end{equation}
In four dimensions when tetrad index run over only four values such relation 
is impossible in general and according to (\ref{metric}) it could be always 
written in multi dimensions.

Let's suppose that in multi-dimensional flat space-time there exists (1+3)-brane 
with arbitrary geometry. In order to simplify demonstration of the idea, let us 
first consider the case of only one extra space-like dimension. Generalization 
for arbitrary dimensions and signature is obvious. 

Let's say that the equation of the branes surface in the Cartesian 5-dimensional 
coordinates $X^A$ has the form:
\begin{equation} \label{surface}
W(X^A) = 0 ~.
\end{equation}
Introducing the function
\begin{equation} \label{xi}
\xi(X^A) = \frac{W(X^A)}{\sqrt{|\partial_BW\partial^BW|}} ~,
\end{equation}
the metric of pseudo-Euclidean bulk (1+4)-space can be transformed to the 
Gaussian normal 
coordinates 
\begin{equation} \label{gaussian}
ds^2 = - d\xi^2 + g_{\alpha\beta}(\xi,x^\nu)dx^\alpha dx^\beta ~.
\end{equation}
Since $\xi = 0 $ is the equation of the hyper-surface, the induced metric 
$g_{\alpha\beta}(0,x^\nu)$, which determines the geometry on the brane, is the 
same 4-dimensional metric as used in (\ref{metric}) for the embedding.

Introducing unit normal vector to the brane
\begin{equation} \label{n}
n^A = \partial^A \xi ~|_{\xi = 0} ~,
\end{equation}
one can decompose tensors of bulk space-time in a standard way (see e. g. 
\cite{MTW}). 

In the Gaussian system of coordinates (\ref{gaussian}) the Christoffel 
symbols on the brane are: 
\begin{equation}\label{Cris}
\Gamma^{\alpha}_{\nu\lambda}=h^{A\alpha}\partial_\lambda h_{A\nu} ~.
\end{equation}
Raising and lowering of Greek indices is made with the induced metric tensor 
$g_{\alpha\beta}$ and Latin indices with 5-dimensional Minkowskian metric 
tensor $\eta_{AB}$. The Christoffel symbols containing two or three indices 
$\xi$ are equal to zero. The connections containing just one index $\xi$ are 
forming outer curvature tensor, which, after using (\ref{tetrad}), can be 
written as:
\begin{equation} \label{K}
K_{\alpha\beta} = n_AD_\beta h^A_\alpha = -\Gamma^\xi_{\alpha\beta}= 
\partial_\alpha\partial_\beta \phi^\xi ~,
\end{equation}
where $D_\beta$ denotes covariant derivatives in Gaussian coordinates 
(\ref{gaussian}) and $\phi^\xi$ is the transversal component of the embedding 
function.

Since bulk 5-dimensional space-time is pseudo-Euclidean, its scalar curvature 
is zero: 
\begin{equation}\label{gauss}
^5R = R + K^2 - K_{\alpha\beta}K^{\alpha\beta} = 0 ~.
\end{equation}
From this relation the 4-dimensional scalar curvature $R$ can be expressed with 
the quadratic combinations of the extrinsic curvature $ K_{\alpha\beta}$. Thus, 
using (\ref{K}) Hilbert's 4-dimensional gravitational action (after removing of 
boundary terms) can be written in the form: 
\begin{equation}\label{GS}
S_{g}=-M_{Pl}^2 \int R \sqrt{-g}d^4x = M_{Pl}^2\int 
\left(\Box \phi^\xi \Box \phi^\xi - \partial_\alpha\partial_\beta \phi^\xi 
\partial^\alpha\partial^\beta \phi^\xi \right) \sqrt{-g}d^4x ~,
\end{equation}
where $\Box = \partial_\alpha\partial^\alpha$ is the 4-dimensional wave operator 
and $g$ is the determinant in Gaussian coordinates. It is clear now that 
embedding theory allows us to rewrite 4-dimensional gravitational action in 
terms of derivatives of the normal components of some multi-dimensional vector.

Now let us consider the bulk massless vector field $A^B$ that obeys 
5-dimensional Maxwell's equations
\begin{equation} \label{Maxwell}
\partial_A F^{AB} = 0 ~,
\end{equation}
where $F_{AB} = \partial_A A_B - \partial_B A_A $ is the ordinary field 
strength. We avoid connection of the functions $A^B$ with the bulk coordinates 
$X^A$, not to restrict ourselves with pure geometrical interpretation. 

The action for Maxwell's field can be written in the general form:
\begin{equation} \label{MS1}
S_A = - \frac{1}{4}\int F_{AB}F^{AB} d^5X = - \frac{1}{2}\int 
[\partial_A A_B (\partial^A A^B + \partial^B A^A) - 2\partial_A 
A^A\partial_B A^B] d^5X ~.
\end{equation}
We shall demonstrate below that on the brane this action can be reduced to the 
4-dimensional gravity action (\ref{GS}). 

Note that so called vacuum gauge fields - the solutions of the equations
\begin{equation} \label{vacuum}
F_{AB} = 0 ~,
\end{equation}
are always present in the space-time. These fields are solutions of Maxwell's 
equations (\ref{Maxwell}) as well. If there are no topological defects in 
space-time, the solutions of (\ref{vacuum}) are pure gauges. The example of 
non-trivial solution of (\ref{vacuum}) in space-time with the linear defect is 
Aharonov-Bohm field ( see e. g. \cite{AB}). For the brane the normal to its 
surface components of vacuum gauge fields also are non-trivial and we shall 
show that they can resemble gravity on the brane. 

In the Gaussian coordinates (\ref{gaussian}) the ansatz which satisfies 
(\ref{vacuum}) and has the symmetries of the brane, with the accuracy of 
constants, can be written in the form:
\begin{equation} \label{ansatz}
A^\alpha = G(\xi)\partial^\alpha \phi (x^\beta ) ~, ~~~~~
A^\xi = \phi (x^\beta) \partial^\xi G(\xi) ~,
\end{equation}
where $G(\xi )$ and $\phi (x^\beta)$ are some functions depending on the fifth 
coordinate $\xi$, and four-coordinates $x^\beta $ respectively. We assume that 
$G(\xi )$ is an even function of $\xi$ and the integrals from $G(\xi )$ and 
from its derivative are convergent. Simple example of such a function is 
$exp(\xi^2)$. So, we choose the ansatz: 
\begin{equation} \label{sol}
A^\alpha = \frac{c}{\epsilon^{3/2}} \partial^\alpha \phi^\xi (x^\beta ) 
~exp \left(-\frac{\xi^2}{\epsilon^2}\right) 
~, ~~~~~
A^\xi = - \frac{2c}{\epsilon^{7/2}}~\xi \phi^\xi (x^\beta)
~exp \left(-\frac{\xi^2}{\epsilon^2}\right) ~,
\end{equation}
where $\epsilon$ is the brane width and $c$ is some dimensionless constant. We 
assume that the width of the brane $\epsilon$ is constant all along the surface. 

Inserting the ansatz (\ref{sol}) into action integral (\ref{MS1}), and 
integrating by the normal coordinate $\xi$ we receive the induced action on 
the brane: 
\begin{equation} \label{MS2} \nonumber
S_A = \sqrt{\frac{\pi}{2}}~ \frac{c^2}{\epsilon^2}
\int \left(\Box\phi^\xi\Box\phi^\xi -
\partial_\alpha\partial_\beta\phi^\xi \partial^\alpha\partial^\beta\phi^\xi 
\right) \sqrt{-g}d^4x ~,
\end{equation}
where summing is made with the intrinsic metric $g_{\alpha\beta}(x^\nu)$. If we 
put
\begin{equation} \label{Plank}
M_{Pl}^2 = \sqrt{\frac{\pi}{2}}~\frac{c^2}{\epsilon^2}~,
\end{equation}
the effective action of 5-dimensional vector field (\ref{MS2}) becomes 
equivalent to Hilbert's action for 4-dimensional gravity (\ref{GS}). So, 
4-dimensional Einstein's equations on the brane can be received from 
Maxwell's multi-dimensional equations in flat space-time.

The same result can be obtained in the case $N > 5$. Now $\xi$ and 
$\phi^\xi$ in (\ref{n}), (\ref{K}) and (\ref{sol}) must be replaced by 
$\xi^i$ and $\phi^i$, and action integrals (\ref{GS}) and (\ref{MS2}) 
transform to the sum:
\begin{equation} \label{SM3}
S_{\Sigma} = M_{Pl}^2\eta_{ij}\int \left(\Box\phi^i\Box\phi^j -
\partial_\alpha\partial_\beta\phi^i \partial^\alpha\partial^\beta\phi^j 
\right) \sqrt{-g}d^4x ~,
\end{equation}
where $\eta_{ij}$ is the Minkowskian metric of the normal space to the brane. 
Small Latin indices $i,j,\dots$ enumerates extra $(N - 4)$ coordinates. If we 
assume that all the brane widths are equal to $\epsilon$ for the scale in 
(\ref{SM3}) we have: 
\begin{equation} \label{M}
M_{Pl}^2 = \frac{c^2}{\epsilon^2} \left(\frac{\pi}{2}\right)^{(N-4)/2} ~.
\end{equation}

Now we want to consider the particular example 6-dimensional pseudo-Euclidean 
space with the signature $(2+4)$ to demonstrate 4-dimensional unification of 
gravity and Electro-Magnetism based on Maxwell's equations. In (\ref{SM3}) 
there exists only derivatives of $\phi^i$ (which we connected with gravity) 
and it is clear that $F_{\alpha\beta}$ is zero. Now we want to insert in 
(\ref{sol}) also the non-integrable part $a^\alpha (x^\beta )$ and identify 
it with the 4-dimensional vector-potentials. 

The (2+4)-space is interesting object for investigations. As it was found a 
long time ago, in this space Schwarzschild's metric admits embedding 
\cite{ros}, and that non-linear 15-parameter conformal group can be written 
as a linear Lorentz-type transformations \cite{Dir}. Another fact is that 
5-dimensional metric with exponential warp factor (needed to cancel 
cosmological constant \cite{GRS}) can be embedded in the Euclidean 
(2+4)-space \cite{Go}. For compact Kaluza-Klein models the (2+4)-space was 
studied in \cite{PI} and in the context of brane models with non-factorizable 
geometry in \cite{GM}.

Let us insert $a^\alpha (x^\beta )$ in (\ref{sol}) and rewrite it in the form:
\begin{eqnarray} \label{sol1}
A^\alpha = \frac{c}{\epsilon^2}~\partial^\alpha \phi^\tau(x^\beta ) 
~exp \left(-\frac{\tau^2}{\epsilon^2}\right) + 
\frac{c}{\epsilon^2}~ \partial^\alpha \phi^\kappa (x^\beta ) 
~exp \left(-\frac{\kappa^2}{\epsilon^2}\right) + a^\alpha 
~exp \left(- \frac{\tau^2+\kappa^2}{\epsilon^2}\right) ~, \nonumber \\
A^\tau = - \frac{2c}{\epsilon^4} ~\tau\phi^\tau (x^\beta)
~exp \left(-\frac{\tau^2}{\epsilon^2}\right) ~, \\
A^\kappa = - \frac{2c}{\epsilon^4} 
~\kappa\phi^\kappa (x^\beta)
~exp \left(-\frac{\kappa^2}{\epsilon^2}\right) ~, 
\nonumber 
\end{eqnarray}
where $\kappa$ and $\tau$ are the normal to the brane space-like and time-like 
coordinates respectively. 

Using the ansatz (\ref{sol1}) Maxwell's action in (2+4)-space takes the form:
\begin{eqnarray} \label{F}
S_A = - \frac{1}{4}\int F_{AB}F^{AB} d^6X = \nonumber \\
= - \frac{\pi c^2}{2\epsilon^2} \int R \sqrt{-g}d^4x -\frac{1}{4\epsilon^2}
\int exp\left[-\frac{2(\tau^2+\kappa^2)}{\epsilon^2}\right]
\left[ f_{\alpha\beta}f^{\alpha\beta} + \frac{4}{\epsilon^4}(\tau^2- 
\kappa^2)a_\alpha a^\alpha \right]\sqrt{-g}d^4xd\tau d\kappa = \\
= - \int \left( M_{Pl}^2 R +\frac{1}{4} f_{\alpha\beta}f^{\alpha\beta} 
\right)\sqrt{-g}d^4x ~,\nonumber
\end{eqnarray}
where $f_{\alpha\beta} = \partial_\alpha a_\beta - \partial_\beta a_\alpha$. 
So, 6-dimensional Maxwell's action on the brane is reduced 
to the 4-dimensional Einstein-Maxwell action. As in \cite{GM} the brane is 
placed at $\tau^2 = \kappa^2$, that means that it moves with the speed of light 
in transversal (1+1)-space.

In Gaussian coordinates (\ref{gaussian}) Maxwell's 6-dimensional equations, 
after inserting there the ansatz (\ref{sol1}), reduces to the system:
\begin{eqnarray}\label{maxwell}
\frac{1}{\sqrt{-g}}\partial_\alpha (\sqrt{-g}f^{\alpha\beta}) - 
\frac{4}{\epsilon^4}(\tau^2 - \kappa^2) a^\beta = 0 ~, \nonumber \\
\partial_\alpha a^\alpha = 0 ~.
\end{eqnarray}
One can see that the situation is similar to superconductors. On the brane 
($\tau^2 = \kappa^2$) photon is massless and obtains the large mass 
$\sim 1/\epsilon $ in the direction normal to the brane.

We want to note that even if we don't want to connect $\phi^i$ with gravity, 
the ansatz (\ref{sol1}) can be used for trapping photons on the brane in the 
(2+4)-space (for localization of massless vector fields on a brane, see 
\cite{GM,vector}).

Hence, in this paper the equivalence of the descriptions of 4-dimensional 
gravity both with the intrinsic metric, and with the multi-dimensional vector 
field was demonstrated. We have not considered here models for the field that 
forms the brane, as well as the question how the brane geometry is changed 
because of couplings with 4-dimensional matter was not raised.
\medskip

{\bf Acknowledgements:} Author would like to acknowledge the hospitality
extended during his visits at the Abdus Salam International Centre for
Theoretical Physics where this work was done.

\end{document}